# Charge carrier injection electroluminescence with CO functionalized tips on single molecular emitters.

Jiří Doležal[1], Pablo Merino [2,3]*, Jesus Redondo[1], Lukáš Ondič[1], Aleš Cahlík [1,4], Martin Švec [1,4]*

[1] Institute of Physics, Czech Academy of Sciences, Praha, Czech Republic
[2] Instituto de Ciencia de Materiales de Madrid, CSIC, Sor Juana Inés de la Cruz 3, E28049, Madrid, Spain.
[3] Instituto de Física Fundamental, CSIC, Serrano 121, E28006, Madrid, Spain.
[4] Regional Center for Advanced Materials and Technologies, Olomouc, Czech Republic

*corresponding authors: svec@fzu.cz, pablo.merino@csic.es

**Abstract:**

We investigate electroluminescence of single molecular emitters on NaCl on Ag(111) and Au(111) with submolecular resolution in a low-temperature scanning probe microscope with tunneling current, atomic force and light detection capabilities. Role of the tip state is studied in the photon maps of a prototypical emitter, zinc phthalocyanine (ZnPc), using metal and CO-metal tips. CO-functionalization is found to have a dramatic impact on the resolution and contrast of the photon maps due to the localized overlap of the p-orbitals on the tip with the molecular orbitals of the emitter. The possibility of using the same CO-functionalized tip for tip-enhanced photon detection and high resolution atomic force is demonstrated. We study the electroluminescence of ZnPc, induced by charge carrier injection at sufficiently high bias voltages. We propose that the distinct level alignment of the ZnPc frontier orbital with the Au(111) and Ag(111) Fermi levels governs the primary excitation mechanisms as the injection of electrons and holes from the tip into the molecule, respectively. These findings put forward the importance of the tip status in the photon maps and contribute to a better understanding of the photophysics of organic molecules on surfaces.

Optical properties of molecules are intimately connected to their atomic and electronic structure. In recent years we have seen spectacular advancement in the study of single molecular emitters at the nanoscale. [1,2] Quantum emission, [3,4] energy transfer processes [5,6] or Raman spectroscopy can now be probed with sub-Å resolution. [7,8] However, the underlying mechanisms involved in the optoelectronic response at the scale of individual molecules have not been fully elucidated yet. Scanning tunneling microscopy-induced luminescence (STML), provides a unique platform to examine charge-to-photon conversion on individual molecules with atomic scale precision. Several excitation mechanisms have been proposed to explain the observed molecular emission under varying tunneling conditions, namely direct charge injection, where pairs of opposite charges meet in the molecular emitter to form singlet [9,10] or triplet states, [11] energy transfer between the plasmon and molecular excitons [6,12] or between different excitonic states, [5] or triplet mediated up-conversion [13]. However, the extent and interplay of these mechanisms are still actively debated.

Independently, advances in the non-contact atomic force microscopy (ncAFM) methodology opened unique way to directly image the atomic structure of individual molecules, which has been already applied to an immense variety of systems. [14] Combining the information from force maps and photon maps is a long-sought goal since it could help to explore new photophysical phenomena on individual molecular emitters with unprecedented level of detail. [15] However, for imaging with ncAFM, the tip needs to be functionalized by a probe particle (most frequently a CO). The impact of the probe particle on a tip apex in local optical spectroscopy is an open question that has not been addressed for the case of STML yet. [16,17,18] The spatial resolution and contrast of the optical signal may be affected, since the exciton formation rate within molecules strongly depends on the charge transport channels, which are defined by the spatial and energetic electronic orbital structure of the tip apex and the molecular interface.[1] As the probe functionalization usually results in a lower junction conductivity, it may prove to be a factor in obtaining sufficiently strong signals on multiple layers of insulating material. Moreover, STML relies on relatively large biases necessary to induce the excitonic state. In contrast, a standard ncAFM imaging typically uses very low biases in order to avoid spurious effects of the electrostatic force, high tunneling currents resulting in a crosstalk between the tunneling and force channels or a loss of the probe particle. Despite recent advances,[15] it is still not clear whether the STML methodology is compatible with the ncAFM framework.

In the present letter we use a combination of STML and nc-AFM to investigate a prototypical example of individual molecular emitters: ZnPc/NaCl on two different metal substrates: Ag(111) and Au(111) (see schematics in Fig.1a). We correlate the photon and force maps of individual emitters and determine the feasibility and role of CO tips as the STML probes. We show that CO functionalization is compatible with electroluminescence measurements in STM and leads to enhanced lateral resolution of photon maps due to the involvement of the p-orbitals of the CO-tip in the tunneling process.[19] Since the ZnPc molecules on NaCl have a natural tendency to alternate between two adsorption geometries when exposed to higher tunneling currents,[20] we characterize the effect of the motion on the force and photon maps by comparison to a stabilized ZnPc molecule. Finally, by correlating photon maps with the spatial distribution of the molecular orbitals, we address the leading excitation mechanism on each inspected substrate.

In Fig.1b we present a STM constant-current (CC) overview image of the system, prepared on the Ag(111) substrate. Individual adsorbed ZnPc molecules deposited onto a cold sample (4K) appear scattered on top of a trilayer NaCl area as well as on the bare metal. Overview shows two ZnPc molecules that exhibit a symmetrical 16-lobe appearance in the CC images scanned at -2.2 V. This appearance is characteristic of their fast angular switching between two equivalent adsorption configurations triggered by the flow of the tunneling current (in the order of 1 pA).[20] If $|V_{BIAS}|$ is decreased sufficiently (<1 V, 6 pA), the switching rate drops well below frequency cutoff of the tunneling current preamplifier of the STM (1 kHz). This can be manifested by a telegraph noise in the tunneling current channel (Fig.1c) with a tip positioned above the molecule lobe at constant height (CH) with the feedback loop open. The overview in Fig.1b also shows one molecule with 8 lobes, likely stabilized by a defect of the NaCl structure underneath. We observed a very similar behavior for ZnPc/NaCl on the Au(111) substrate, confirming previous work.[21] Thus, at the biases needed for inducing electroluminescence ($|V_{BIAS}| > 1.8$ V), the most commonly found state of the molecules is dynamic.

By controlled nanoindentation of the tip into a clean patch of the substrate metal, we have been able to obtain tips that had a strong optical response (measured by plasmon intensity at a fixed current), and had been successfully functionalized by CO picked up from the NaCl layers (see Supp. Info.). With these tips we were able to record luminescence and high-resolution AFM/STM signals. The important consequence of the CO attachment is the sensitization of the probe apex to lateral forces stemming from the Pauli repulsion between the oxygen and the atoms of the molecular emitter, detected as a change of the sensor resonant frequency. [22] This frequency provides a high submolecular resolution, closely related to the atomic structure of the emitter.[23][24]

To evaluate the impact of the ZnPc rotation on each measured channel, including the photon emission, we measured two sets of maps on: (i) a rotating and (ii) a stabilized molecule with a CO-functionalized Au tip. The spatially-resolved maps of the tunnel current, frequency shift and photon channels for both molecules were taken within the same measurement session. The data shown in the upper row of Fig.2 originate from a ZnPc, that has been stabilized in one adsorption position, by moving it to the vicinity of a step edge of the NaCl trilayer. At a bias of +2.2 V, the CC STM of this molecule essentially reproduces the shape of the ZnPc LUMO with its characteristic 8 lobes and a pronounced central part.[20] However, in the CH mode, a triple squarelike pattern emerges in the molecular center, which is a telltale sign of imaging by a mixed s- and p-wave tip, confirming the presence of the CO at the apex.[19] The frequency shift map, taken at a considerably lower voltage (25 mV) and with the tip closer to the surface by 250 pm, resolves the pyrrole backbone of the ZnPc molecule and its peripheral benzene rings with high spatial precision. The rotating molecule shown in the bottom row of Fig.2 exhibits the previously described STM contrast of 16 lobes. [25][26][5] The AFM taken at 25 mV confirms that there are two distinguishable angular positions, since the angular switching slows down enough to hold each position at timescale of minutes.

We have recorded the electroluminescence spectra and photon maps at $V_{bias}$=+2.2 V on ZnPc in both states. The spectral fingerprints of the rotating and the stable molecule do not exhibit significant differences within our resolution; the intense main fluorescence line is located at 653 nm (1.9 eV), accompanied by red-shifted vibronic sideband (Fig 1b inset). The spatially resolved photon maps of the integrated main emission line (640-660 nm) of the rotating and fixed ZnPc molecules are shown in the right part of Fig.2. For both rotating and stabilized molecule, these photon maps resemble at first glance their corresponding CH current images. This observation points to a direct charge carrier injection from the tip as the main mechanism for excitation that precedes the luminescence at this voltage. [27][28][11][13] A deeper inspection of the maps reveals a dip in the photon intensity at the center of the molecule, although the density of states and the CC image reaches a maximum at the same spot. The existence of a minimum in metallo-phthalocyanine photon maps has been observed previously with metallic tips and has been attributed to the spatial variation of the exciton coupling to the nanocavity plasmon modes. [29][30][26]

To better understand the impact of the tip state on luminescence maps, we have performed a detailed characterization of the ZnPc molecules also with bare metal tips. In the case of Au tips, both the CC and CH STM modes show a very similar type of contrast (upper row of Fig. 3), contrary to the data obtained by Au CO-tips. This rather blunt contrast at the periphery and in the center of the molecule is a consequence of two factors: the overall s-wave character of the bare Au tip and smearing of the features by the rapid angular switching. The photon map, again, closely resembles the CH tunneling current maps, except for a dip near to the center. We have observed this dip at various off-center positions (see Supp. Info.), whose lateral shift seems to be related to the mesoscopic tip shape. This is not surprising, as the real tips are likely to deviate from the spherical shape and therefore they are expected to screen asymmetrically the transient dipole moments involved in the emission process. [30]

Detailed luminescence spectra, normalized by the current, taken at various positions above the molecule (Fig. 3) with both CO and metallic tips confirm the strong intensity modulation of the photon maps by the tunneling current and suppression of the signal near the molecular center. Within our spectral resolution, the overall shape of the spectra does not vary neither at the main line nor at the vibrational sideband. This observation reveals the direct effect of the tip frontier orbitals in combination with the shape of the ZnPc LUMO on the electroluminescence maps and indicates that injection of electrons from the tip to the molecule triggers the excitation, that can eventually lead to a luminescent event. We now raise the question whether an analogous mechanism is also valid for biases where hole injection is the primary mode of charge transport. Using negative biases on the ZnPc/NaCl/Au(111) system is not suitable since it leads to ZnPc transiently switching to a cationic state. This would result in the preferential activation of a different excitonic state, manifested by a lower energy photon emission below the detection threshold of our optical setup.[21] However, it has been reported that using Ag(111) as a metallic substrate leads to the same 1.9 eV exciton radiative decay detectable at negative biases. [25]

Therefore, we performed experiment on the ZnPc/NaCl/Ag(111) system and obtain photon maps with both metal (Ag-coated Au tip) and CO-functionalized tips. Here, the photon maps are also closely related to the respective tunneling current. The current flows predominantly through the HOMO of individual rotating molecules decoupled from the substrate by trilayer of NaCl. The effect of the CO on the STM images, apart from a sharpening of the contrast of the peripheral benzenes, is the appearance of the central pyrrolic ring. This pyrrolic ring is very pronounced in the CO-tip maps, but absent from the metal-tip images. This feature is a good indicator for the difference between the current and photon maps in the central part of the molecule. For the metal tip, the photon map faithfully copies the current except for a small central area, where a dip occurs similarly as in the positive biases on Au substrate. For the CO-tip, although the general contrasts of the photon and STM maps are similar, the photon signature of the pyrrolic ring is weaker with respect to the periphery of the molecule, contrary to the corresponding CH STM.

Comparing the photon maps obtained on the Au(111) at positive biases and Ag(111) at negative biases with metal and CO-tips, we see striking differences corresponding mainly to the distinct spatial modulation of the charge carrier transport. On both substrates the individual normalized spectra show an identical lineshape, which implies that we observe the radiative decay of the same final $S_1$ excited state. This general correspondence between the tunneling current and the luminescence channels indicates that the fluorescence is triggered by a resonant tunnelling charge injection into the molecular orbital. This brings us to the discussion of how different excitation mechanisms occurring under given circumstances (bias polarity, workfunction of the substrate) lead to the same excited state of ZnPc and its subsequent radiative decay.

The alignments of the molecular levels relative to $E_F$ of the substrate are remarkably different between Au and Ag, as it can be inferred from the differential conductance curves of each system, caused by the higher work function of Au(111) compared to Ag(111). [25][21] On trilayer NaCl/Au(111) substrate, the ZnPc LUMO is detected at +2.2 eV and HOMO at -1.1 eV. [21] On the Ag(111) substrate, the situation is LUMO at +1.0 eV and HOMO at -2.3 eV. [25] Possible band bending due to an electric field of the tip leading to a diode-like excitation process has been a matter of debate in the literature recently, with controversial conclusions. The absence of a measurable Stark effect indicates a negligible effect of the electric field on the excitonic state.[31] Importantly, the predominant 1.9 eV radiative exciton generation occurs in opposite polarities on Au and Ag substrate, and scales with the total electric current transported through the ZnPc LUMO and HOMO levels, respectively. We cannot entirely rule out an energy transfer mechanism between ballistically transported inelastic electrons, decaying in the substrate, and the exciton, however important details as the bipolar character of the electroluminescence is not captured in such a scheme. [28][13][6]

We propose two alternative explanations for the observed phenomena, involving either an opposite charge carrier capture or an energy conversion process within the transiently charged ZnPc. In Fig. 4a,f the schemes of the charge injection and capture process are shown for the two substrates with their corresponding energy level diagrams in Fig. 4b,d,e,f. In particular, the mechanism for luminescence generation for the ZnPc molecules deposited on Au(111) inspected at $V_{bias}$= +2.2V would be the following. Electrons are injected from the tip into the LUMO of the molecule. Most of the electrons injected from the tip into the molecule will continue tunnelling through the NaCl contributing to the net tunnel current, with very low residence times on the molecule, until one electron remains long enough to attract a hole from the substrate through the NaCl into the transiently charged ZnPc- anion. The energy levels of the excited molecule renormalize and a bound electron-hole $S_1$ exciton forms which may couple to the nanocavity plasmon and become detected. This mechanism is only possible when the LUMO energy above the substrate $E_F$ is near or larger than the exciton energy; otherwise the energy for the $S_1$ generation is insufficient.[11] This would explain the absence of any significant photon signal with Ag substrate at positive biases. A converse mechanism in the system on Ag(111) will lead to the same exciton, if it is initiated by a hole injection from the tip into the HOMO of the molecule at negative biases (Fig.4e) and followed by an electron capture from the substrate. The HOMO of the ZnPc/NaCl/Ag(111) separation from the $E_F$ is sufficiently large to induce the $S_1$ exciton. In the second proposed type of mechanism (Fig.4c, 4f), the charge capture would not occur; instead a decay of the transient charge would induce the $S_1$ exciton by an energy transfer within the molecule. In both proposed mechanisms, the contrast of the photon maps will be reflecting the spatial modulation of the corresponding charge carrier transport routes, corresponding to injection of electrons on Au(111) and holes on Ag(111). The precise internal process that links the charge injection to the exciton generation and the photon signal dependence on the nanocavity electrical field and tunneling current have yet to be elucidated. More detailed STML measurements together with time-resolved STML may help to enlighten this point. [3, 27]

To conclude, we have performed a real-space multi-channel study of the prototypical molecular emitter, ZnPc, by combining tunneling current, force, and photon maps of the quantum emitter on Au(111) and Ag(111). We demonstrated compatibility of the CO-functionalized tips with electroluminescence measurements. STML maps were compared for two states of the ZnPc - the dynamical angularly-switching state, induced by bias and current and a state stabilized in a single position. The CO functionalization leads to the enhancement of the spatial resolution of photon maps as compared to bare metal tips. By studying the spectra and photon maps obtained by injecting electrons and holes into the molecule by metal and CO-tips, we have found that photon intensity maps are intimately linked to the spatial modulation of the tunneling current, corresponding to charge transport from the tip to the molecule. Electron injection from the tip into the molecule LUMO is triggering the exciton formation events in the ZnPc on NaCl/Au(111) while hole injection into its HOMO initiates the process at NaCl/Ag(111) substrate. Our findings identify the role of the tip status in the STML photon map contrast and open new avenues to spatially-resolved optical measurements of single-molecule emitters in combination with submolecular-resolution AFM.

**Associated content:**

*Supporting Information*: The Supporting Information is available free of charge on the ACS Publications website at DOI: XX.XXXX/acs.nanolett.XXXXXXX.

Experimental details, plasmonic spectrum of the tips used in the experiments, tip state and intensity minimum in the photon maps and photon maps for a ZnPc dimer. (PDF)

**Author information:**

*Corresponding Author*s
 *E-mail: svec@fzu.cz, pablo.merino@csic.es.

*ORCID*
Pablo Merino: 0000-0002-0267-4020
Martin Svec: 0000-0003-0369-8144

*Notes*
The authors declare no competing financial interests.

**Aknowledgements:**

P. M thanks the ERC Synergy Program (grant no. ERC-2013-SYG-610256, Nanocosmos) for financial support and the "Comunidad de Madrid" for its support to the FotoArt-CM Project (S2018/NMT-4367) through the Program of R&D activities between research groups in Technologies 2013, co-financed by European Structural Funds. MS, JD and JR acknowledge the Czech grant agency funding no. 17-24210Y.

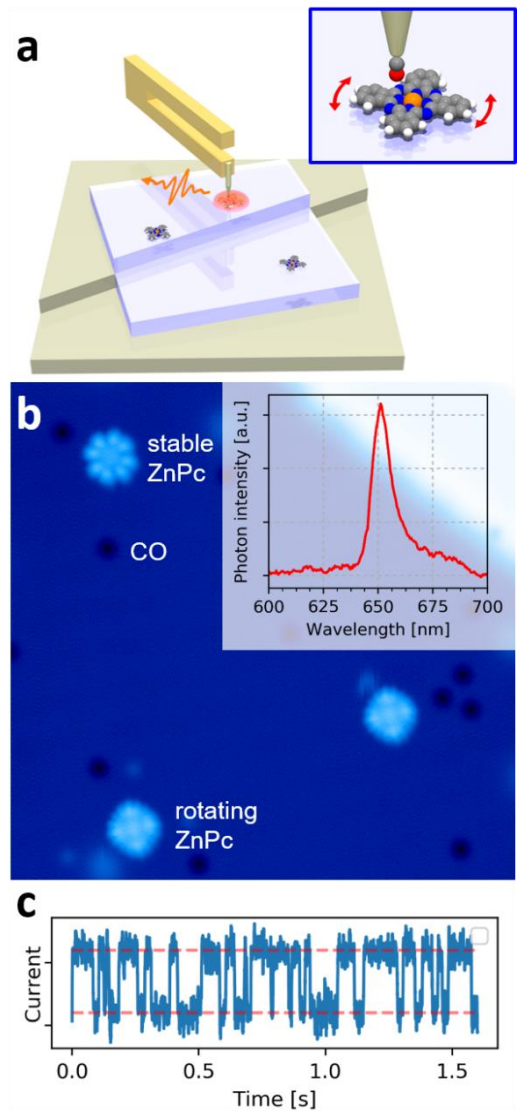

*Figure 1*: a) Schematics of the experimental set-up, in which the STM and ncAFM signals are collected together with electroluminescence signal on single ZnPc molecules, electronically decoupled from a metal substrate by an insulating NaCl layer. For this purpose, a CO-functionalized tip is used (shown in the inset). Applied current/bias leads to a configuration switching of the adsorption position of the molecule (inset). b) STM constant-current image of rotating and stable ZnPc molecules on the surface of trilayer NaCl on Ag(111) surface, with an example luminescence spectrum in the inset. The 23 x 23 nm² image was taken at -2.2 V, 5 pA. c) Telegraphic noise recorded on a rotating ZnPc at 1.0 V, 6 pA.

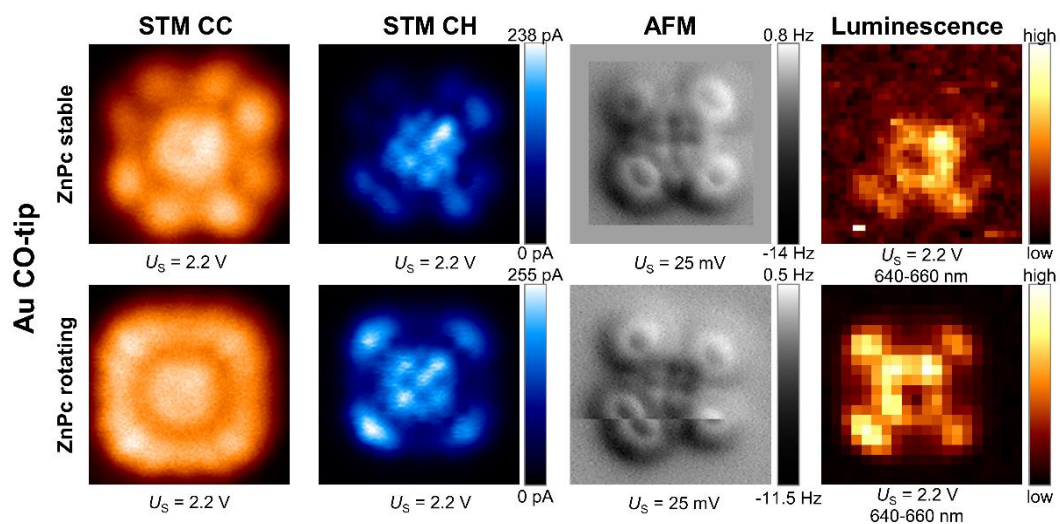

*Figure 2:* Comparison of spatially resolved constant current (CC) STM, constant height (CH) STM , AFM, and photon images, obtained by Au-CO-tips on a stable (top row) and a rotating ZnPc molecule (bottom row), adsorbed on trilayer NaCl on Au(111). All image sizes (including padding at the bottom AFM image) were 2.2 x 2.2 nm². The constant current STM images were taken with 1 pA setpoint.

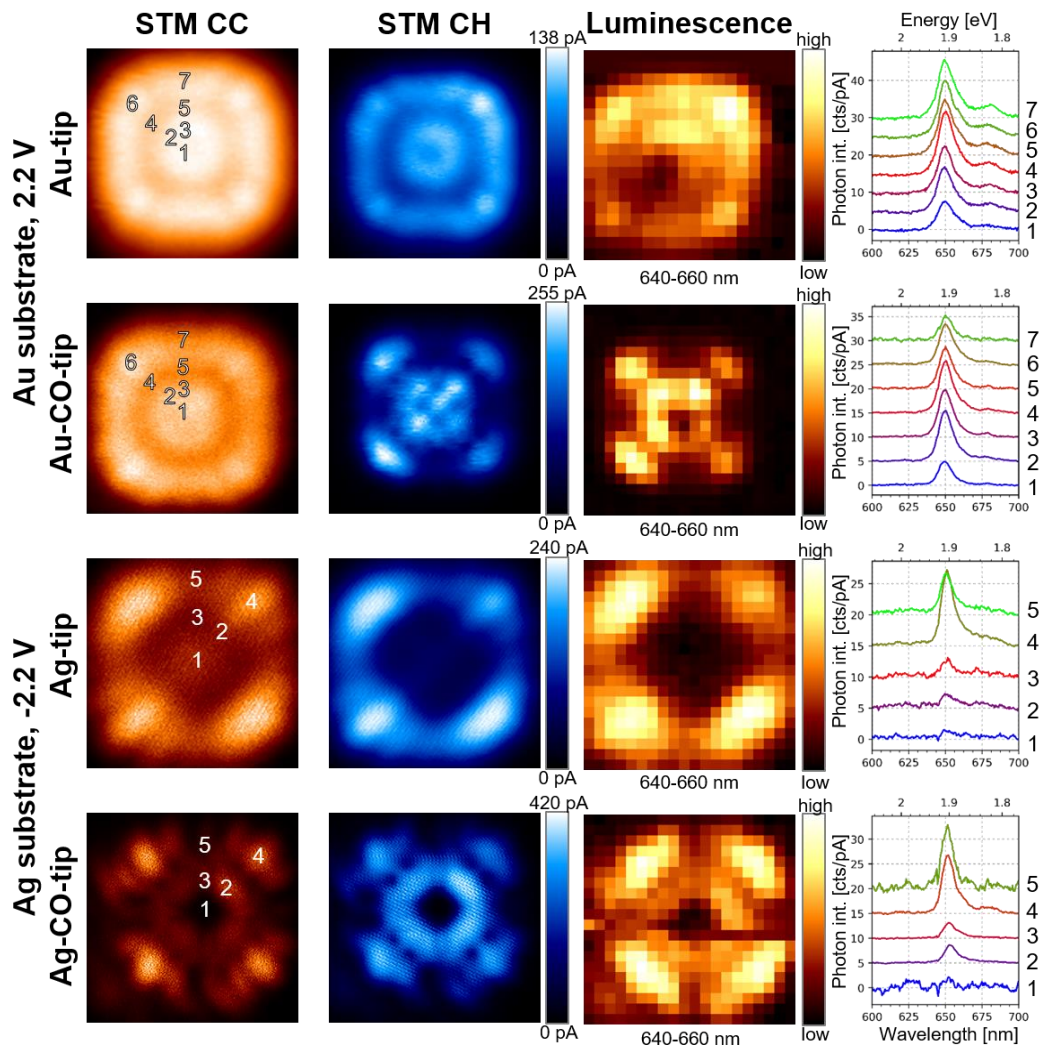

*Figure 3*: Effect of the functionalization (CO-tips vs. bare-metal tips) and substrate on the photon maps and spectra of the dynamic ZnPc molecules. The two upper rows show data taken above ZnPc on the 3ML-NaCl/Au(111). The two bottom rows are results obtained above ZnPc on 3ML-NaCl/Ag(111). STM images at constant current (CC), constant height (CH), STM-induced luminescence (STML) maps at constant height measured with metal and CO tips are presented for comparison. The numbered spectra shown on the right-hand side column have been taken at the positions marked in the CC STM images. All image sizes are 2.2 x 2.2 nm², tunneling current setpoints for CC images were 10 pA and 1 pA on Au and Ag substrate, respectively.

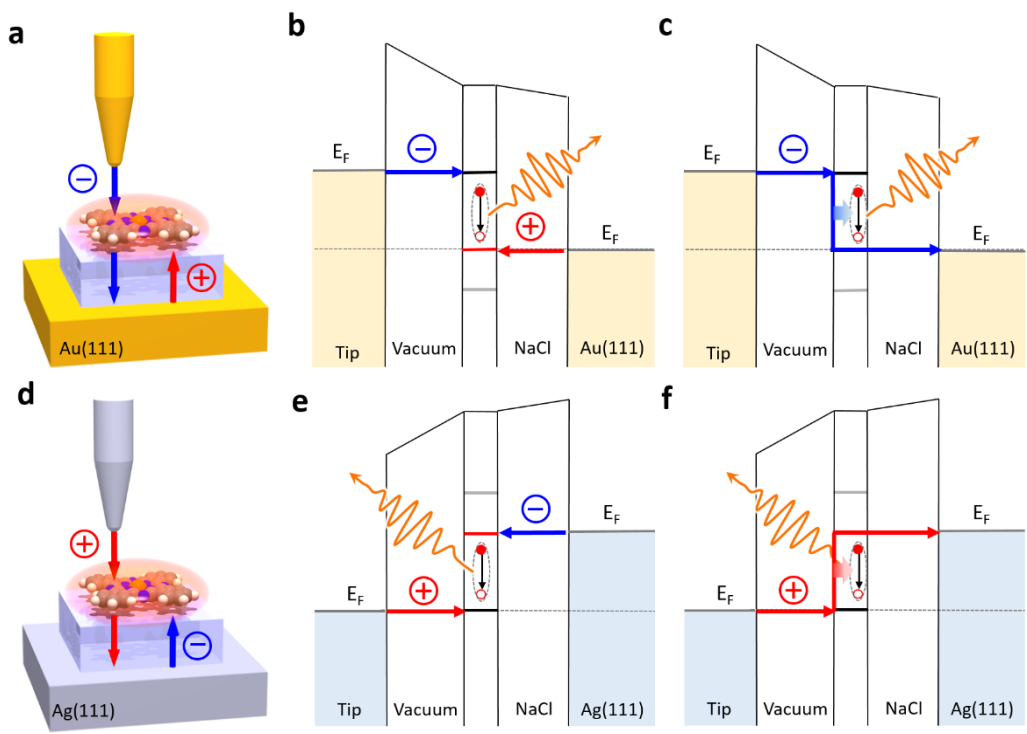

**Figure 4**: a) *Graphical representation and b) and c) band schemes of the proposed excitation processes of ZnPc on 3ML-NaCl/Au(111) in which an electron injection from the tip into the molecule through the tunneling barrier (blue arrows) is the primary mechanism leading to radiant exciton formation. d) and e) f) corresponding graphical representation and the band scheme for the excitation processes on 3ML-NaCl/Ag(111) primarily induced by hole injection into the molecule form the tip.*

**Table of contents (TOC)**

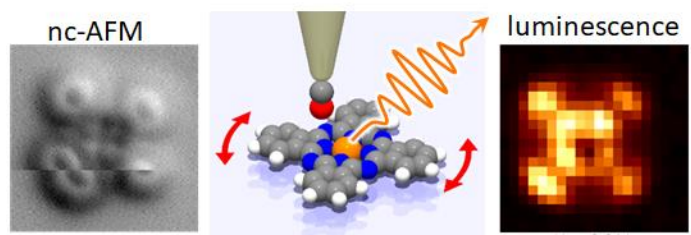

**References.**

1. Kuhnke, K., Große, C., Merino, P., Kern, K. Atomic-Scale Imaging and Spectroscopy of Electroluminescence at Molecular Interfaces. *Chem. Rev.* **117,** 5174-5222 (2017).

2. Rossel, F., Pivetta, M., Schneider, W.-D. Luminescence Experiments on Supported Molecules with the Scanning Tunneling Microscope. *Surf. Sci. Reports* **65,** 129–144 (2010).

3. Merino, P., Große, C., Rosławska, A., Kuhnke, K., Kern, K. Exciton dynamics of C60-based single-photon emitters explored by Hanbury Brown–Twiss scanning tunnelling microscopy. *Nat. Commun.* **6,** 8461 (2015).

4. Luo, Y. *et al.* Electrically Driven Single-Photon Superradiance from Molecular Chains in a Plasmonic Nanocavity. *Phys. Rev. Lett.* **122,** 233901 (2019).

5. Imada, H. *et al.* Real-space investigation of energy transfer in heterogeneous molecular dimers. *Nature* **538,** 364–367 (2016).

6. Imada, H. *et al.* Single-Molecule Investigation of Energy Dynamics in a Coupled Plasmon-Exciton System. *Phys. Rev. Lett.* **119,** 13901 (2017).

7. Zhang, R. *et al.* Chemical mapping of a single molecule by plasmon-enhanced Raman scattering. *Nature* **498,** 82–86 (2013).

8. Lee, J., Crampton, K. T., Tallarida, N., Apkarian, V. A. Visualizing vibrational normal modes of a single molecule with atomically confined light. *Nature* **568,** 78–82 (2019).

9. Große, C., Gunnarsson, O., Merino, P., Kuhnke, K., Kern, K. Nanoscale Imaging of Charge Carrier and Exciton Trapping at Structural Defects in Organic Semiconductors. *Nano Lett.* **16,** 2084–2089 (2016).

10. Große, C. *et al.* Submolecular Electroluminescence Mapping of Organic Semiconductors. *ACS Nano* **11,** 1230-1237 (2017).

11. Kimura, K. *et al.* Selective triplet exciton formation in a single molecule. *Nature* **570,** 210–213 (2019).

12. Kröger, J., Doppagne, B., Scheurer, F., Schull, G. Fano Description of Single-Hydrocarbon Fluorescence Excited by a Scanning Tunneling Microscope. *Nano Lett.* **18,** 3407–3413 (2018).

13. Chen, G. *et al.* Spin-Triplet-Mediated Up-Conversion and Crossover Behavior in Single-Molecule Electroluminescence. *Phys. Rev. Lett.* **122,** 177401 (2019).

14. Gross, L. *et al.* Atomic Force Microscopy for Molecular Structure Elucidation. *Angew. Chemie Int. Ed.* **57,** 3888–3908 (2018).

15. Kaiser, K., Gross, L., Schulz, F. A Single-Molecule Chemical Reaction Studied by High-Resolution Atomic Force Microscopy and Scanning Tunneling Microscopy Induced Light Emission. *ACS Nano* **13,** 6947-6954 (2019). doi:10.1021/acsnano.9b01852

16. Tallarida, N., Lee, J., Apkarian, V. A. Tip-Enhanced Raman Spectromicroscopy on the Angstrom Scale: Bare and CO-Terminated Ag Tips. *ACS Nano* **11,** 11393–11401 (2017).

17. Gieseking, R. L. M., Lee, J., Tallarida, N., Apkarian, V. A., Schatz, G. C. Bias-Dependent Chemical Enhancement and Nonclassical Stark Effect in Tip-Enhanced Raman Spectromicroscopy of CO-Terminated Ag Tips. *J. Phys. Chem. Lett.* **9,** 3074–3080 (2018).

18. Lee, J., Tallarida, N., Chen, X., Jensen, L., Apkarian, V. A. Microscopy with a single-molecule scanning electrometer. *Sci. Adv.* **4,** eaat5472 (2018).

19. Gross, L. *et al.* High-Resolution Molecular Orbital Imaging Using a $p$-Wave STM Tip. *Phys. Rev. Lett.* **107,** 86101 (2011).

20. Miwa, K., Imada, H., Kawahara, S., Kim, Y. Effects of molecule-insulator interaction on geometric property of a single phthalocyanine molecule adsorbed on an ultrathin NaCl film. *Phys. Rev. B* **93,** 165419 (2016).

21. Doppagne, B. *et al.* Electrofluorochromism at the single-molecule level. *Science* **361,** 251 LP-255 (2018).

22. de la Torre, B. *et al.* Submolecular Resolution by Variation of the Inelastic Electron Tunneling Spectroscopy Amplitude and its Relation to the AFM/STM Signal. *Phys. Rev. Lett.* **119,** 166001 (2017).

23. Hapala, P. *et al.* Mechanism of high-resolution STM/AFM imaging with functionalized tips. *Phys. Rev. B* **90,** 85421 (2014).

24. Moll, N., Gross, L., Mohn, F., Curioni, A., Meyer, G. The mechanisms underlying the enhanced resolution of atomic force microscopy with functionalized tips. *New J. Phys.* **12,** 125020 (2010).

25. Doppagne, B. *et al.* Vibronic Spectroscopy with Submolecular Resolution from STM-Induced Electroluminescence. *Phys. Rev. Lett.* **118,** 127401 (2017).

26. Zhang, Y. *et al.* Visualizing coherent intermolecular dipole–dipole coupling in real space. *Nature* **531,** 623–627 (2016).

27. Rosławska, A. *et al.* Single Charge and Exciton Dynamics Probed by Molecular-Scale-Induced Electroluminescence. *Nano Lett.* **18,** 4001–4007 (2018).

28. Merino, P. *et al.* Bimodal exciton-plasmon light sources controlled by local charge carrier injection. *Sci. Adv.* **4,** eaap8349 (2018).

29. Chen, C., Chu, P., Bobisch, C. A., Mills, D. L. & Ho, W. Viewing the Interior of a Single Molecule: Vibronically Resolved Photon Imaging at Submolecular Resolution. *Phys. Rev. Lett.* **105,** 217402 (2010).

30. Neuman, T., Esteban, R., Casanova, D., García-Vidal, F. J. & Aizpurua, J. Coupling of Molecular Emitters and Plasmonic Cavities beyond the Point-Dipole Approximation. *Nano Lett.* **18,** 2358–2364 (2018).

31. Kuhnke, K. *et al.* Pentacene Excitons in Strong Electric Fields. *ChemPhysChem* **19,** 277–283 (2018).